# What Do AI-Generated Images Want?

Amanda Wasielewski

Department of ALM
Uppsala University
amanda.wasielewski@abm.uu.se

**Abstract:**
W.J.T. Mitchell's influential essay "What do pictures want?" shifts the theoretical focus away from the interpretative act of understanding pictures and from the motivations of the humans who create them to the possibility that the picture itself is an entity with agency and wants. In this article, I reframe Mitchell's question in light of contemporary AI image generation tools to ask: what do AI-generated images want? Drawing from art historical discourse on the nature of abstraction, I argue that AI-generated images want specificity and concreteness because they are fundamentally abstract. Multimodal text-to-image models, which are the primary subject of this article, are based on the premise that text and image are interchangeable or exchangeable tokens and that there is a commensurability between them, at least as represented mathematically in data. The user pipeline that sees textual input become visual output, however, obscures this representational regress and makes it seem like one form transforms into the other—as if by magic.

**Keywords:** generative AI, W.J.T. Mitchell, image studies, art history, abstraction, divinity

In March 2023 an open letter calling for a temporary stop to AI development was published online and quickly drew media attention.[1] It was signed by high-profile figures in the tech industry, including those who were instrumental in the development of deep learning such as Yoshua Bengio and Yann LeCun. In addressing the dangers of AI, however, the letter was not primarily focused on current failings, such as the perpetuation of bias and stereotype, the effect on labor, the spread of misinformation, the fact that commercial models use copyrighted material without clearance, or environmental impact. Instead, the primary rationale these AI acolytes proposed for pausing AI development was that AI would soon have enough autonomy and intelligence to destroy humanity. While this may seem like a step beyond rational—and very real—fears around AI, the stature of the signatories attest to the fact that an eschatological angst has gripped Silicon Valley. For OpenAI, a company founded by tech figures deeply involved in AI mysticism, the desire to push generative AI into all aspects of life has, ironically, been accompanied by the extreme existential dread expressed in the open letter.[2]

---

[1] "Pause Giant AI Experiments: An Open Letter," Future of Life Institute, March 22, 2023, https://futureoflife.org/open-letter/pause-giant-ai-experiments/; Kari Paul, "Letter Signed by Elon Musk Demanding AI Research Pause Sparks Controversy," Technology, *The Guardian*, April 1, 2023, https://www.theguardian.com/technology/2023/mar/31/ai-research-pause-elon-musk-chatgpt; Cade Metz and Gregory Schmidt, "Elon Musk and Others Call for Pause on A.I., Citing 'Profound Risks to Society,'" Technology, *The New York Times*, March 29, 2023, https://www.nytimes.com/2023/03/29/technology/ai-artificial-intelligence-musk-risks.html; Samantha Murphy Kelly, "Elon Musk and Other Tech Leaders Call for Pause in 'out of Control' AI Race | CNN Business," CNN, March 29, 2023, https://www.cnn.com/2023/03/29/tech/ai-letter-elon-musk-tech-leaders.

In order to make sense of the connections among the various belief systems that contribute to this spiritual movement in the tech industry, Timnit Gebru and Émile P. Torres created the acronym TESCREAL, which stands for the tech cults of Transhumanism, Extropianism, Singularitarianism, (modern) Cosmism, Rationalist ideology, Effective Altruism, and Longtermism. Some of these groups optimistically look forward to the melding of human and machine while others fear the powers of autonomous artificial intelligence agents. However, they share a worldview that has tipped from speculation to dogmatic faith. Gebru and Torres argue that these movements are rooted in the history of eugenics and are premised on a belief that so-called AGI (Artificial General Intelligence) may bring about the end of human civilization. They are thus based on reactionary ideas of white superiority that have been instrumental in justifying western subjugation of the world through empire, colonialism and slavery.

Members of these groups often subscribe to the idea that generative AI models, such as large language models, are achieving consciousness or beginning to express autonomous desires. If the idea that AI has—or will soon have—consciousness is taken seriously, one might naturally wonder what it wants. I am not, however, among those concerned with the speculative idea of AI consciousness. I am far more concerned about the current and very real harms the AI industry creates. So, rather than take this question seriously in the sense that TESCREAL believers do, I would like to instead reframe it in terms of the agency W.J.T. Mitchell grants pictures in his seminal essay, "What Do Pictures Want?" In short, I address how AI-generated images may be thought to have latent desire akin to Mitchell's pictures. Both Mitchell's discussion of the power of pictures and contemporary discussions around AI speak to the human impulse to attribute beliefs and projections of imagination to human creations.

Unlike Mitchell's question, which positions pictures as desiring objects, my own question calls for no radical reframing of its subject as something agentic or, even desiring. Ascribing agency to generative AI is, as the aforementioned cults surround it attest, already a fait accompli, whether one believes that AGI (or the annihilation of humanity) is around the corner or not. The architects of contemporary AI models, drawing from science fiction, have already positioned their tools as something potentially agentic. In light of this, I revisit Mitchell's discussion of what pictures want and what this means for the relationship between language and image for AI-generated images. If generative AI is going to usurp humanity in its present form (that is, as large language models and image-producing models), it will do so through an exchange of representational forms. Words and pictures have never been so potent.

In his essay, Mitchell ultimately avoids providing an overarching answer to the question of what pictures want, as he is more concerned with pointing to the affective subjectivity of pictures than explaining their individual motivations. I intend, however, to put forth an answer to my own version of this question. What do AI-generated images want? AI-generated images want specificity. They want concreteness. More precisely, they want naturalistic representation and they want to create a semblance of reality. Why do they want this? Because they are, in contrast to these desires, fundamentally abstract. AI-generated images are abstractions that desire an escape from the vast representational gulf between their source and eventual appearance. Multimodal generative AI models, which are the primary subject of this article rather than image-only generative models like GAN, are based on the premise that text and image are interchangeable or exchangeable tokens and that there is a commensurability between them, at least as represented mathematically in data forms.[3] The user pipeline that sees textual input become visual output, however, obscures this representational regress and makes it seem like one form transforms into the other—as if by magic.

---

[2] Karen Hao, *Empire of AI* (Penguin Press, 2025).

**Potent Pictures**

Mitchell's essay and its evocative title have had many afterlives or, at least, a long run of influence.[4] Not only did Mitchell himself publish several versions of the essay but other scholars have taken up the question in a variety of forms.[5] The essay began its life as a conference presentation in October 1994.[6] There were then three versions published in 1996, 1997, and 2005 that I reference in the present text. The first version in print was a revision of the original essay, which, confusingly, first appeared in 1997 *after* the revision. Published in the journal *October* in 1996, "What do pictures *really* want?" states that it is a "condensed and rather different version" than the version that would appear in the 1997 anthology edited by Terry Smith on modernism and masculinity in visual culture.[7] This anthology version—which I hazard to call the 'original' version—was published a year later with the title "What Do Pictures Want? An Idea of Visual Culture."[8] The third publication I cite has largely been treated as the definitive version in subsequent scholarship, since it gives its name to an eponymous book: *What Do Pictures Want? The Lives and Loves of Images* published in 2005. It also includes a coda with "Frequently Asked Questions," an unusual move for an academic text that points not only to the enduring influence of the essay but the confusion over whether it presented a workable framework for visual analysis.

---

[3] GAN stands for Generative Adversarial Network. See Ian Goodfellow et al., "Generative Adversarial Nets," in *Advances in Neural Information Processing Systems*, ed. Z. Ghahramani et al., vol. 27 (Curran Associates, Inc., 2014).

[4] The anthology version of the essay reveals that the question arose in a review of Mitchell's book in the *Village Voice* that declared Mitchell's book *Picture Theory* should have been called "What Do Images Want?" instead. W.J.T. Mitchell, "What Do Pictures Want? An Idea of Visual Culture," in In Visible Touch: Modernism and Masculinity, ed. Terry Smith (University of Chicago Press, 1997), 215.

[5] A sampling of these include J. Hillis Miller, "What Do Stories about Pictures Want?," *Critical Inquiry* 34, no. S2 (2008): S59–97, https://doi.org/10.1086/529090; Jacques Rancière, "Do Pictures Really Want to Live?," *Culture, Theory and Critique* 50, nos. 2–3 (2009): 123–32, https://doi.org/10.1080/14735780903240083; Chris Gosden, "What Do Objects Want?," *Journal of Archaeological Method and Theory* 12, no. 3 (2005): 193–211, https://doi.org/10.1007/s10816-005-6928-x.

[6] Mitchell, "What Do Pictures Want? An Idea of Visual Culture," 215.

[7] W. J. T. Mitchell, "What Do Pictures 'Really' Want?," *October* 77 (1996): 71–82, https://doi.org/10.2307/778960.

[8] Mitchell, "What Do Pictures Want? An Idea of Visual Culture."

Although the central argument remains more or less intact between these versions of the essay, the differences show some nuances in Mitchell's thought that I highlight below. Before delving into Mitchell's more esoteric arguments, however, I wanted to highlight the fact that the condensed versions contain certain differences in the callousness with which race and racial discrimination is treated in relation to the hypothetical subjectivity of pictures. I would argue that all versions of the text are problematic in their conflation of the image, positioned as an affective person, with the desires of human beings who are oppressed and subjugated through gender/racial discrimination and colonialist domination. However, the original version of the essay makes the infractions of the subsequent versions seem subtle by comparison. The most troubling language found in the anthology version seems to have been edited and smoothed out of subsequent versions.[9] In the *October* revision, Mitchell points to the potential (already voiced?) objection to this framing by acknowledging that some might find it a "tasteless appropriation."[10] This does not excuse the conceit, but it seems to have mitigated its impact.

Beyond these problematic points of comparison, the primary aim of Mitchell's essay, as noted, is to shift the theoretical focus away from the interpretative act of understanding pictures and from the motivations of the human who create them to the possibility that the picture itself is an entity with agency and wants. He thus does not specifically answer the question of what pictures want but, rather, reframes the object of inquiry around the picture as an entity that has desires. He defines pictures not as images alone but as a construct similar to a holy trinity: one unity that is a "three-way intersection" of image, object, and discourse.[11] In this article I have chosen to label AI-generated visual output 'images' and pose the central question as 'images' rather than 'pictures' because AI-generated visual output lacks one of the essential components of Mitchell's trinity: objecthood. I will return to this point shortly.

The mystical power of pictures is, in fact, an important component of Mitchell's argument and is cited in its various guises as magic, totemism, fetishism, idolatry and animism. Despite modern attempts to pull away from the irrationality of belief systems that imbue pictures with agency and life, Mitchell argues, "we are stuck with our magical, premodern attitudes toward objects, especially pictures, and our task is not to overcome these attitudes but to understand them."[12] He cites the example of the commercial advertisement as a powerful type of modern picture, though it is generally not considered religious or mystical in the way that, for example, the golden calf of the biblical Exodus is (which is also cited by Mitchell). Advertising imagery has an inherent power in its implementation that is not quite within the grasp or understanding of those who put it out in the world nor is its power over those who see it something that can be rationalized.

---

[9] In a particularly egregious example, which is part of a reading of Frantz Fanon, Mitchell asks, "Are pictures n*****s?" Even if this question was posed within a good faith effort to incorporate theories of race, gender, and postcolonialism, it nevertheless trivializes these discourses. I hesitate to even quote this here, as it seems an academic question best left to its present state of obscurity, but it is so absurdly insensitive that I felt it needed to be mentioned, given that I am framing my own argument around this essay and the assumption of its continued relevance to discussions of images, text, and media today. Subsequent versions of Mitchell's essay excise this passage quite conspicuously. See Mitchell, "What Do Pictures Want? An Idea of Visual Culture," 219.
[10] Mitchell, "What Do Pictures 'Really' Want?," 71.
[11] Mitchell, "What Do Pictures Want? An Idea of Visual Culture," 217.
[12] Mitchell, "What Do Pictures 'Really' Want?," 72.

The thing that fosters this mysterious power of images is, according to Mitchell, premised on an "act of forgetting." He writes, "…the worshipper projects agency, consciousness, and subjectivity into an inanimate image, and then forgets the act of projection, mistaking the 'life' that appears in the image as something autonomous and self-generated."[13] That humans can be thus startled by their own creations is compounded when those creations are automated. The act of forgetting for generative AI is readily apparent, and the people who have designed these tools are the most keen to forget. They create, thereby, the conditions for their subsequent movement toward iconoclasm.

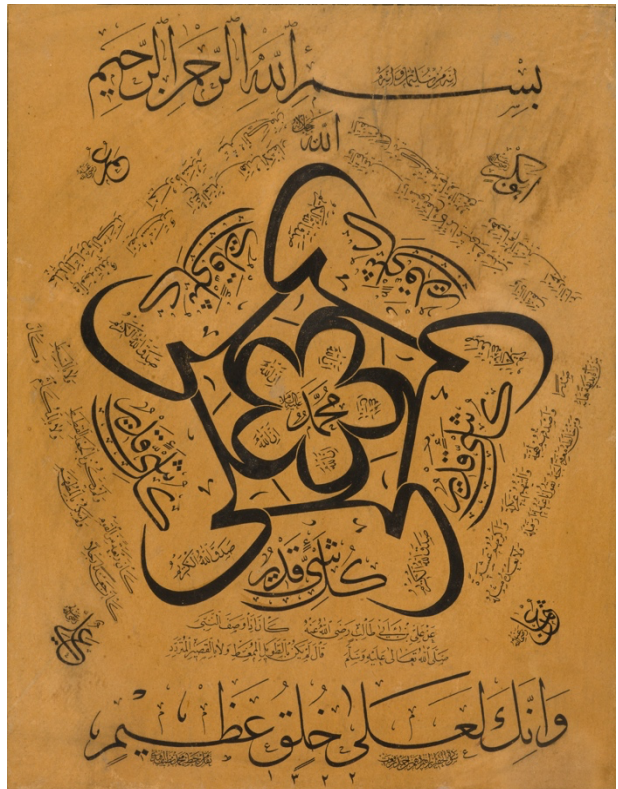

*Figure 1* Hilyah, 1322 AH (1904-5 AD), Ottoman Turkey, ink on coated paper mounted on board, 32.3 x 25 cm, Nasser D. Khalili Collection of Islamic Art, Credit: Khalili Collections / CC-BY-SA 3.0 IGO

A discussion of the potency of the image in isolation, however, does a disservice to its ancient religious foil: text. Iconoclasm, bans on images, and the avoidance of pictures in religious practice have historically led deeper into an equally potent projection of agency on the textual representation of divinity. The Protestant practice of Christianity, for example, thus rejects the icons of the saints or the Virgin Mary that are ubiquitous in Catholicism to move closer to the supposed source of divinity in another representative form, that is, the "word of God." Sacred texts defy the positioning of text as merely descriptive or instrumentalized in the service of recording. Islamic art has also historically treated text in this way. For example, an Ottoman *hilyah* [Fig. 1] uses verses of the Qur'an in a highly stylized and decorative pattern of writing, emphasizing both the pictorial form of the script and the potency of the message as an icon wrapped into one. It illustrates the difficulty in truly separating representative forms. As such, it is an explicit example of the synthesis of image and text that Mitchell calls the "imagetext."[14] Scholars of Islamic art have, of course, long sought to understand the relationship of script and pictorial representation across the centuries of art production inflected by Islamic religious practice.[15]

---

[13] Mitchell, "What Do Pictures Want? An Idea of Visual Culture," 225.

One of Mitchell's enduring arguments, which flows through multiple works on text and image including the essay discussed here, is that there are no purely visual media.[16] The intermingling of visual and verbal forms of representation in media does not, however, mean that they are "formally commensurable," according to Mitchell.[17] Pictures and language are, in this framing, not interchangeable or directly exchangeable, as much as scholars of visual culture might try to wrangle them together or place them in a relation with one another. They are not pure forms or unitary tokens to be exchanged or translated one into the other. This raises the question: what does it mean that AI-generated images are produced by a model that does not semantically separate the data forms representing text and image? How do these abstract mathematical relations, in turn, produce concrete forms?

In his earlier works such as *Iconology* (1986) and *Picture Theory* (1994), Mitchell ruminated on the question: "What is an image?"[18] Discussing the shift that has occurred in the post-AI landscape, media scholar Jussi Parikka asks, "So what is an image if it is lost somewhere inside the machinations of photography and datasets and machine learning?"[19] Somewhere between Mitchell and Parikka, the image got lost. Perhaps a better question to pose, then, is not *what* is an image but *where* is an image? Is the digital code that constitutes it and the trained model that metabolizes or 'learns' from it still the image? The continual reconstitution of images via data makes finding a locus and, thus, a concrete reality for images more and more difficult.[20] "Some decades ago," Parikka continues, "scholars could still respond by saying that images are like language."[21] What has changed, then? The power of the AI-generated image is mixed up not only in the magical or idolatrous power of images that Mitchell references but also in its concrete relationship to text mathematically represented by digital tools.

Whereas Mitchell's mixed media could primarily be thought of as mixed due to its discursive resonances, AI-generated images are *formally* mixed between image and language. Even if they are functionally different, they are mathematically enmeshed. For CLIP (contrastive language image pre-training), the model that underwrites the functioning of text-to-image AI tools, these data forms exist within a single model space. This is the trick that allows a user to type in a prompt to produce an image but it is also the trick that complicates the status of these representational and symbolic forms, datafied as they are within the latent space of the model.[22] Mitchell's conclusion in his essay was: "Pictures want equal rights with language, not to be turned into language."[23] The statement points to the traditional subservience of imagery to explanation, description, and ekphrasis. Has this order been now overturned and have images been liberated? Or, in light of text-to-image models, could we turn this statement around and ask: does text wants equal rights with pictures, not to be turned into pictures?

---

[14] W. J. T. Mitchell, *Picture Theory: Essays on Verbal and Visual Representation* (University of Chicago Press, 1995), 89, n.9.
[15] Christiane Gruber, ed., *The Image Debate: Figural Representation in Islam and Across the World* (Gingko, 2019).
[16] Mitchell, *Picture Theory*, 5; W. J.T. Mitchell, "There Are No Visual Media," *Journal of Visual Culture* 4, no. 2 (2005): 257–66, https://doi.org/10.1177/1470412905054673.
[17] Mitchell, *Picture Theory*, 5.
[18] W.J.T. Mitchell, *Iconology: Image, Text, Ideology* (University of Chicago Press, 1986); Mitchell, *Picture Theory*.
[19] Jussi Parikka, *Operational Images: From the Visual to the Invisual* (University of Minnesota Press, 2023), 93.
[20] See, for example, Offert's discussion of the constitution of scientific illustrations and images via the use of GANs (generative adversarial networks) in Fabian Offert, "Latent Deep Space: Generative Adversarial Networks (GANs) in the Sciences," *Media+Environment* 3, no. 2 (2021), https://doi.org/10.1525/001c.29905.
[21] Parikka, *Operational Images: From the Visual to the Invisual*, 93.

**Words in want of a picture**
Debates in the twentieth century regarding semiotics and the nature of visual, textual, and verbal expressions have been numerous, and Mitchell was central in many of the later debates on the topic in the anglophone sphere.[24] Specifically, Mitchell claimed that, at the close of the century, the academic trend that Richard Rorty had dubbed "the linguistic turn" had shifted into a "pictorial turn."[25] Without wishing to get too deep into the larger scope of these debates, I want to focus here on the specific case of how Mitchell's writing might inform an understanding of text-to-image models and the digital images generated from these tools.

If, at some point, text and image could be thought of as two different ways of recording expressions or thoughts externally, with no shared basis other than those processes internal to the human mind, then they could subsequently be conceived as something that goes through a process of translation, one to the other. The pivot upon which a mountain of philosophy and theory on the text-image relationship rests, therefore, is the conversion between the two—or, otherwise, the incommensurability between the two. However, digital text and images are programmatic appearances based on something that is, at least on the surface, commensurable: numbers and their mathematical relationships.[26] Into this scenario of representational regress comes the AI models based on CLIP.

To address this paradigm shift, I need to take one step back in the history of digital information creation and organization. Before the recent apotheosis of artificial neural networks, big data and the database were the dominant element of digital culture that scholars sought to theorize.[27] In this context, media scholar Sean Cubitt adapted Mitchell's elastic formulation to ask, "What do databases want?" He does so in order to understand the "mass image," a term he uses for the accumulation of big image data after social media, surveillance and other digital tools that have produced superlative imagery. In defining this term, Cubitt points to the unknowability of the contents of big image data due to scale, which is out of proportion with human-level knowledge comprehension.

---

[22] Alec Radford et al., "Learning Transferable Visual Models From Natural Language Supervision," arXiv:2103.00020, preprint, arXiv, February 26, 2021, https://doi.org/10.48550/arXiv.2103.00020.
[23] W. J. T. Mitchell, *What Do Pictures Want?: The Lives and Loves of Images* (University of Chicago Press, 2005), 47.
[24] In these discussions, including those of Mitchell, the role of French theory, particularly the writing of Michel Foucault and Jacques Derrida, has been central. See Mitchell, *Picture Theory*, 83–107.
[25] Mitchell, *Picture Theory*, 11–34.
[26] Peter Galison describes debates in mathematics regarding the nature of mathematical representation, see Peter Galison, "Images Scatter into Data, Data Gather into Images," in *Iconoclash: Beyond the Image Wars in Science, Religion, and Art*, ed. Bruno Latour and Peter Weibel (MIT Press, 2002), 303.
[27] Danah Boyd and Kate Crawford, "Critical Questions for Big Data: Provocations for a Cultural, Technological, and Scholarly Phenomenon," *Information, Communication & Society* 15, no. 5 (2012): 662–79; Kate Crawford et al., "Critiquing Big Data: Politics, Ethics, Epistemology," *International Journal of Communication* 8, no. 0 (2014): 0; Rob Kitchin, *The Data Revolution: Big Data, Open Data, Data Infrastructures and Their Consequences* (SAGE Publications Ltd, 2014).

Referencing the turn from the still image to the moving image of cinema, he writes that the still image sought a closer relationship to reality via the accumulated image of cinema and that the mass image represents a step further in this direction. He argues:

> Instead of trying to produce a more perfect image, it [the mass image] presents the total agglutination of every possible image in such a way as to supplant the history that the image was alienated from in the first instance. This is how the mass image not only reconstructs the world in its own image but offers itself in the place of reality. […] Photography's original project was to turn the world and ourselves into a picture in order to understand it. Mass photography reverses the order: we need the mass image in order to understand—and reconstruct—ourselves.[28]

And if big data presented a problem for the human subject in the face of the scale of this mass image, artificial neural networks provided the way out in some sense. AI image machines create the singular out of this multiplicity: an individual image from the mass of big image data.

Thus, the simple reason why neural networks and machine learning have exploded in use over the past decade is that they make large-scale data useable. However, the use for big data in AI has quickly shifted in recent years from organizing existing media to producing new media. In order to do so, an even higher order of superlative data was needed, i.e., a webscrape of just about all existing online data was necessary to train the largest language models and image models in use today. In sum, big data needed artificial neural networks to make sense of it but, with the turn toward generative AI, the neural networks needed even bigger data. Multimodality, the training of different data forms such as text and image into a single model space, was thus positioned as the Rosetta stone of AI. To truly be useful, generative AI needed models for formal translation.

A need is not a want, however. Looking for the deeper desire behind a text-to-image translation tool, I turn to the publicity statements for some of the platforms that use this single model space for visual creation. Stability AI claims, for example, that "Stable Diffusion is a deep learning model used for converting text to images. It can generate high-quality, photo-realistic images that look like real photographs by simply inputting any text."[29] The idea here is not only a process of translation but a translation to something "real"—or at least with the appearance of reality via photography. This is a revealing statement in that the first port of call seems to be the formation of a semblance of reality. Another publicity statement, this one from OpenAI as regards their tool DALL·E, states, "DALL·E 2 can create original, realistic images and art from a text description. It can combine concepts, attributes, and styles."[30] Again, "realistic" is front and center. Concepts, attributes, and style thus serve the purpose of instantiating reality (or some form that serves in the business of creating a reality).

---

[28] Sean Cubitt, "The Uncertainty of the Mass Image: Logistics and Behaviours," in *The Uncertain Image*, ed. Ulrik Ekman et al. (Routledge, 2019), 19.
[29] "Stable Diffusion Online," accessed August 4, 2025, https://stablediffusionweb.com/#faq.
[30] "DALL·E 2," November 3, 2022, https://openai.com/index/dall-e-2/.

To return to the central question of this article, then, what do AI-generated images want? If the marketing of these tools are to be believed, they want to turn words into reality—and into objects. Visual generative AI tools build, in part, on computer vision systems designed for object recognition. An image for object recognition is understood as something that merely comprises the objects depicted in it and, thus, a generated image is the reverse: a collection of objects instantiated or textual abstractions objectified. These are, however, phantom objects because they have no individual reference object that they depict.

In Mitchell's formulation, pictures are not just images but the combined material unit of the image-bearing object (the aforementioned holy trinity of image, object, and discourse), and the object-form is the core to Mitchell's claim of pictures' personhood. He writes, "The desires of pictures are not just those of signs signalling other signs, but of bodies calling to bodies."[31] AI-generated images, following this, are the ghosts of signs seeking bodies. They are statistical constructs, i.e., representations of data that are themselves representations of signs. These, in turn, are representations of concepts *en masse*. These distant abstractions of collective thought in various layers of representative form are then pushed out the end of the AI sausage grinder in a seemingly singular unit.

There are no objects and there is no real, in the Baudrilliardian sense, behind these "appearances."[32] It is, of course, already clear to a public inundated with deepfakes that visual AI tools do not produce reality, as the companies promoting these tools would have hoped, but rather "semblances of reality."[33] Modern science relies on statistics to present a picture of what is happening or what is true. The collection of data and its weight toward a certain answer to a research question is the primary way scientists make sense of phenomena that are too large or too complex for simple observation and, so, necessitate copious data collection and processing. Nevertheless, within every large-scale medical study, for example, that makes claims based on statistical extrapolation, there will always be individual case studies to investigate. Visual generative AI has dispensed with cases in favor of a network of signs without traceable origins. What AI-generated images want, therefore, is Mitchell's sense of objecthood for pictures. The object-form is the closest they can come to the real.

**Pictures that don't want to be pictures**
In Mitchell's theorization, there is an unresolved conflation between the desires of the maker of the picture, what/who it represents, and the picture itself as an object. He acknowledges that, in the examples given, there is a "slippage" between "the desire attributed to the represented figure in the picture […] and the picture itself, its materiality as 'support' for a visual image."[34] Mitchell justifies this by arguing that a picture is, in fact, an object and thus has more than a "soul" but also a "selfhood."[35] This assertion is, however, significantly muddied by the specific works he discusses since they all contain representative imagery and many of them, such as the propaganda/recruitment posters he discusses, have simple agendas. Thus, how does one read desire into abstract pictures, given this conflation between a picture's materiality and its conventional iconography?

---

[31] Mitchell, "What Do Pictures Want? An Idea of Visual Culture," 221.
[32] Jean Baudrillard, *Simulacra and Simulation*, trans. Sheila Faria Glaser (University of Michigan Press, 2014), 1.
[33] For longer discussion of Baudrillard in relation to generative AI, see Amanda Wasielewski, "The Latent Objective World: Photography and the Real after Generative AI," in *Virtual Photography: Artificial Intelligence, in-Game, and Extended Reality*, ed. Ali Shobieri and Helen Westgeest (Trascript Verlag, 2024).
[34] Mitchell, "What Do Pictures Want? An Idea of Visual Culture," 221.
[35] Mitchell, "What Do Pictures Want? An Idea of Visual Culture," 221.

The claim that abstract paintings are "pictures that want not to be pictures" appears in all three versions of Mitchell's essay in various forms.[36] The later versions of the essay treat this subject a bit more briefly, so in what follows I refer to the 1997 version. Mitchell's main arguments draw from the work of art historian Michael Fried. Fried is known for his polemical pairing of "absorption" and "theatricality", first articulated in a critical essay denouncing the work of a group of artists' whose three-dimensional, abstract forms in the late 1960s and '70s would later come to be called Minimalism.[37] Fried subsequently spent the majority of his career seeking out this conceptual pairing in different artistic periods, one of which was the latter half of the eighteenth century.[38]

Mitchell draws from Fried's 1988 book on this subject to argue that the desire of pictures is to "take the beholder into the picture" or "the collapse of the distinction between pictures and persons."[39] He asserts that, "The very special sort of pictures that enthrall Fried get what they want by seeming not to want anything, having everything they need – presentness and grace."[40] Fried's absorption thus becomes, in Mitchell's description, a total subsummation. Or, in different terms, Fried's love story becomes Mitchell's horror film. In his discussion of Fried, Mitchell points an implicit thread in his body of work, arguing that Fried's writing on the eighteenth and nineteenth century essentially exists in order to point to the culmination of absorption in the advent of abstract painting.[41]

Mitchell writes, "Abstract paintings, we might say, are pictures that want not to be pictures, that want to be loved and admired for themselves and not for what they represent."[42] Underpinning this assertion is an implicit understanding of abstraction in painting—or works Fried characterizes as absorptive—as those that are solely or primarily concerned with manipulating form to the exclusion of all else. This falls neatly in line with midcentury modern formalist theory spearheaded by Fried's mentor Clement Greenberg.[43] However, it represents a somewhat inaccurate description of the history and development of abstraction in western painting. The spiritual and the formal within western abstraction were always intimately intertwined.

Abstract art in the early twentieth century—or "non-objective art" as it was more commonly known then—has been subject to two competing narratives in the subsequent decades. On the one side, the artists now renowned for their purely abstract compositions, including Piet Mondrian, Hilma af Klint, and Wassily Kandinsky, espoused a spiritualist Theosophy-based rationale for their work and its abstract or non-representational presentation that placed art as an expression of an invisible or otherwise knowable spiritual realm beyond.[44] On the other side, however, art critics claimed abstraction was part of a progressive modern art history, based on an evolution of aesthetic (i.e., sensual) experience of works of art.[45] In the telling of art history, which is pervasive within museums and art history classes alike, it has traditionally been the art critical apparatus that won the day. Caroline Jones describes the modernist mindset as a turn toward making the process of artists' work visible, which is a "suppression (or, if you will, an iconoclasm) of claims to mimetic representation."[46] Only in recent times have the spiritualist underpinnings of this work come back into focus in art historical scholarship and the art market.[47] But why were they ever ignored to begin with?

---

[36] Mitchell, "What Do Pictures 'Really' Want?," 79–81; Mitchell, *What Do Pictures Want?*, 44.
[37] Michael Fried, "Art and Objecthood," *Artforum*, June 1, 1967, https://www.artforum.com/features/art-and-objecthood-211317/.
[38] It should be noted that Fried vehemently denies that there is this direct relationship between his criticism and art historical writing. He does so mostly on the basis that the former is making judgements on the value of the work in question that the latter does not. I do not find these protestations convincing, however. See Michael Fried, *Art and Objecthood: Essays and Reviews* (University of Chicago Press, 1998), 51–52.
[39] Mitchell, "What Do Pictures Want? An Idea of Visual Culture," 220–21.
[40] Mitchell, "What Do Pictures Want? An Idea of Visual Culture," 222.
[41] Fried contends that Mitchell has misread his work and disapproves of Mitchell's summation of his argument on the grounds that Mitchell treats him as a "knowing aficionado instead of a hard-working art historian." See Fried, *Art and Objecthood: Essays and Reviews*, 72–73, n.75.
[42] Mitchell, "What Do Pictures Want? An Idea of Visual Culture," 222–23.
[43] Clement Greenberg, "Post Painterly Abstraction," *Art International* 8, nos. 5–6 (1964): 63.

Indeed, art historians and critics have long chosen to elide or disregard the spiritualist representation found in abstract or non-objective compositions in favor of a stripped-down sense of artistic purity and a materialist understanding of its function. This served the modernist narrative of *art pour l'art* and, later, the Greenbergian narrative of "truth to materials."[48] In his breakthrough essay "Avant-Garde and Kitsch" (1939), Greenberg sets the stage for his dogmatic doctrine of purity, writing, "It has been in search of the absolute that the avant-garde has arrived at 'abstract' or 'nonobjective' art." He goes on to describe this development as part of the human impulse to play God. Rather than a Pygmalion-like artist creating life through representational form, however, modernist abstraction was meant to fulfil the desire to *actually* make something real by rejecting representation. It is the same 'real' that Mitchell's pictures exhibit, except instead of false idols they are nature itself. Greenberg continues:

> The avant-garde poet or artist tries in effect to imitate God by creating something valid solely on its own terms, in the way nature is valid, in the way a landscape – not its picture – is aesthetically valid; something *given*, increate, independent of meanings, similars or originals. Content is to be dissolved so completely into form that the work of art or literature cannot be reduced in whole or in part to anything not itself.[49]

---

[44] This is articulated by the early proponent of non-objective art, including Hilla Rebay and Wassily Kandinsky, in the early twentieth century but it also persisted into the mid-century in American Abstract Expressionism. See Martha Barratt, "Hilma Af Klint: Paintings for the Future Solomon R. Guggenheim Museum, New York 12th October 2018?23rd April," *The Burlington Magazine* 161, no. 1393 (2019): 320–23; Bennett Simon, "Mondrian's Search for Geometric Purity: Creativity and Fixation," *American Imago* 70, no. 3 (2013): 515–55; Mike King, "Concerning the Spiritual in Twentieth-Century Art and Science," *Leonardo* 31, no. 1 (1998): 21–31, https://doi.org/10.2307/1576543; Reinhold Heller, "Kandinsky and Traditions Apocalyptic," *Art Journal* 43, no. 1 (1983): 19–26, https://doi.org/10.2307/776628; Peter Fingesten, "Spirituality, Mysticism and Non-Objective Art," *Art Journal* 21, no. 1 (1961): 2–6, https://doi.org/10.2307/774289.

[45] Proponents of this line of thinking are numerous but, in addition to Greenberg and Fried, key critics in the twentieth century include Clive Bell, Roger Fry, and—much later—Hilton Kramer.

[46] Caroline Jones, "Making Abstraction," in *Iconoclash: Beyond the Image Wars in Science, Religion, and Art*, ed. Bruno Latour and Peter Weibel (MIT Press, 2002), 412.

[47] The case of Hilma af Klint presents a poignant example. So deep and explicit was the spiritual motivation for her work that it hidden from view and thus unacknowledged for many decades. However, after significant art world and art market interest, her descendants have subsequently fought to remove her work from these contexts, which they see as unsuitable to its intent or legacy. They continue, at time of writing, to fight for its removal from the public sphere to a temple-like setting. See Josefin Sköld et al., "Hilma af Klint's art could be hidden from public view—in a temple," Kultur, *Dagens Nyheter* (Sweden), March 10, 2025, https://www.dn.se/kultur/hilma-af-klint-s-art-could-be-hidden-from-public-view-in-a-temple/.

[48] Greenberg, "Post Painterly Abstraction." See also Jones, "Making Abstraction," 412.

And so, a motivation is written onto abstraction that has both nothing and precisely *everything* to do with the spiritualist inclinations or practices of many of the artists who pioneered such work in the western context. It is an inherently spiritual understanding of aesthetics that nevertheless rejects the explicit spirituality of the artists involved.

Greenberg's contemporary Hilla Rebay, an artist and the first director of the Solomon R. Guggenheim Museum, was among those to explicitly tie the givenness Greenberg describes in abstract art to the spiritual, writing in 1936, "The artists of non-objectivity paint with the religious spirit of intuitive creations. […] Art is like the sun, the moon, the rain or the growth of a flower; once it is here it is final and exists in spite of all likes and dislikes."[50] She argues, in a similar manner to artist Wassily Kandinsky, that a turn away from "materialism" was of primary interest to abstract artists of the early twentieth century.[51]

According to art historian David Summers, the art critical impulse to focus on purity of form amounts to iconoclasm. In his ambitious attempt to re-write global art history and the precepts around which western art historians have framed it, Summers addresses some of the core modernist concepts of art, including formalism. He writes, "The aesthetic-formalist view of art underlying abstraction is, however, broader than abstraction, and is in itself iconoclastic in the straightforward sense of ignoring images even when they are present; aesthetic appreciation requires that we *always* look through and around whatever images we find in the art of any tradition to their more essential 'forms' and their relations."[52] As the above statement by Mitchell makes clear, he believes that abstract pictures want not to be pictures because they carry a notion of *art pour l'art* ("want to be loved and admired for themselves and not for what they represent"), but, as Summers argues, they cannot escape representation.

To return to the case of AI-generated images in light of this, then, I would argue that they cannot be defined primarily as singular images, as they appear, but rather as crystallizations of the larger body of data they arise from. They are single images abstracted in the learning process and the data derived from them is then oriented toward particular locations in the latent space of the AI model used to create them but they are, at the same time, inherently collective rather than singular. This is an important distinction because each AI-generated image is undoubtedly singular but, in that singularity, it is still an abstraction of the data. AI-generated images are, in this way, mathematical representations that want to be pictures, if pictures are understood here in the sense of Mitchell's holy trinity (image, object, discourse). They are non-pictures that desire the concreteness of pictures. They seek the objecthood they lack.

---

[49] Clement Greenberg, "Avant-Garde and Kitsch," in *Art in Theory, 1900-2000: An Anthology of Changing Ideas*, ed. Charles Harrison (Blackwell Publishing, 2003), 541.
[50] Hilla Rebay, "A Definition of Non-Objective Painting," *Design* 38, no. 2 (1936): 16.
[51] Wassily Kandinsky, "Concerning the Spiritual in Art," in *Art in Theory, 1900-2000: An Anthology of Changing Ideas*, ed. Charles Harrison (Blackwell Publishing, 2003), 84–85.
[52] David Summers, *Real Spaces: World Art History and the Rise of Western Modernism* (Phaidon, 2003), 253.

Discussing the creation and use of images and image-making techniques in a variety of scientific fields that rely on mathematical representations, historian Peter Galison meditates on precisely this tension between the abstract and concrete. He describes the polemical debates over the use of images in science, writing that images concretize and therefore clarify the complexity of scientific knowledge on one hand but, on the other hand, they deceive due to exactly this self-same specificity. He explores how science—perhaps not unlike society writ large—is "locked in a whirling embrace of iconoclasm and iconophilia."[53]

Essential to this discussion is a conception of *mathematical* abstraction as opposed to visual abstraction. These two concepts of abstraction are expressed in different forms but they nevertheless share partisans who see them as the vehicles of truth or reality. Galison writes, "Abstraction, rigorous abstraction, is exactly that which does not depend on pictures. Abstraction properly conducted proceeds through the formal, the logical, and the systematic."[54] As he shows, however, these two are held in a pattern fluctuating between representation and contra-representation, particularly after the advent of digitization. He writes, "…the impulse to draw the world in its particularity never seems to be able to shed itself of the impulse to abstract, and that search for abstraction is forever pulling back into the material-particular."[55] This ouroboros of the abstract and the specific is evident for AI-generated images too.

As both mathematical—if not visual—abstractions, AI-generated images are essentially abstract. They are the products of the high dimensionality of the model's latent space made concrete. For example, CLIP has a 512-dimensional embedding space and thus requires some type of projection if one were to attempt to visualize this "space." This means that one would have to create a lower order representation so that human perception might grasp the space visually.[56] Projections are, therefore, also an example of what Galison describes as "the haunting oscillation between the concrete and abstract."[57] However, following this argument, AI-generated images are also very much concrete. They exist in this tension. The products of visual generative AI—these statistical abstractions—are thus longing for their opposite: the concrete. And they find it in the concrete form of singular image output while nevertheless remaining essentially abstract.

Each AI-generated image produced, however, cannot avoid the bleed of statistical multiciplicty—of the weight of the mass of image data—from entering an otherwise concrete instantiation, particularly when the weight of collective representation strongly favors a particular outcome. In this way, AI-generated images can never realize the objecthood they aspire to. This can be illustrated with two examples, both of which are images generated with the text-to-image tool Midjourney.

The first image, created from the prompt "a school classroom with a hotel bed in the middle" [Fig. 2] seeks to fulfil one of the key early promises of multimodal image creation: that disparate or unusual pairings of objects may be depicted in a single scene. What one sees in this image, however, is that the bed depicted in the school classroom appears in a way that would suggest a child's bed rather than the neutrally decorated beds found in most hotel rooms. The sheets on the bed in this image are bright blue with a colorfully patterned overlay and multi-color pillows, suggesting a child's bed. The pink and yellow curtain would be more at home in a child's room than in the average hotel room, which usually has a more muted color palette.

---

[53] Galison, "Images Scatter into Data, Data Gather into Images," 301.
[54] Galison, "Images Scatter into Data, Data Gather into Images," 300.
[55] Galison, "Images Scatter into Data, Data Gather into Images," 302.
[56] Alec Radford et al., "Learning Transferable Visual Models From Natural Language Supervision," *Proceedings of the 38th International Conference on Machine Learning*, PMLR, July 2021, 8751.
[57] Galison, "Images Scatter into Data, Data Gather into Images," 323.

In this case, the abstraction of child-like associations in a school classroom effectively bleeds into the image, prohibiting the realization of an objecthood that is singular. That singularity, the hotel bed, is thus polluted by collective associations tied to the other objects associated with the prompt. Hybrid objects such as this are the hallmark of AI-generated images, though they are sometimes well-disguised. They are the products of over-represented associations—in this case, childhood—that are so statistically dominant that it is difficult to avoid having them enter into the elements of the image that, based on the description, should be free from them.

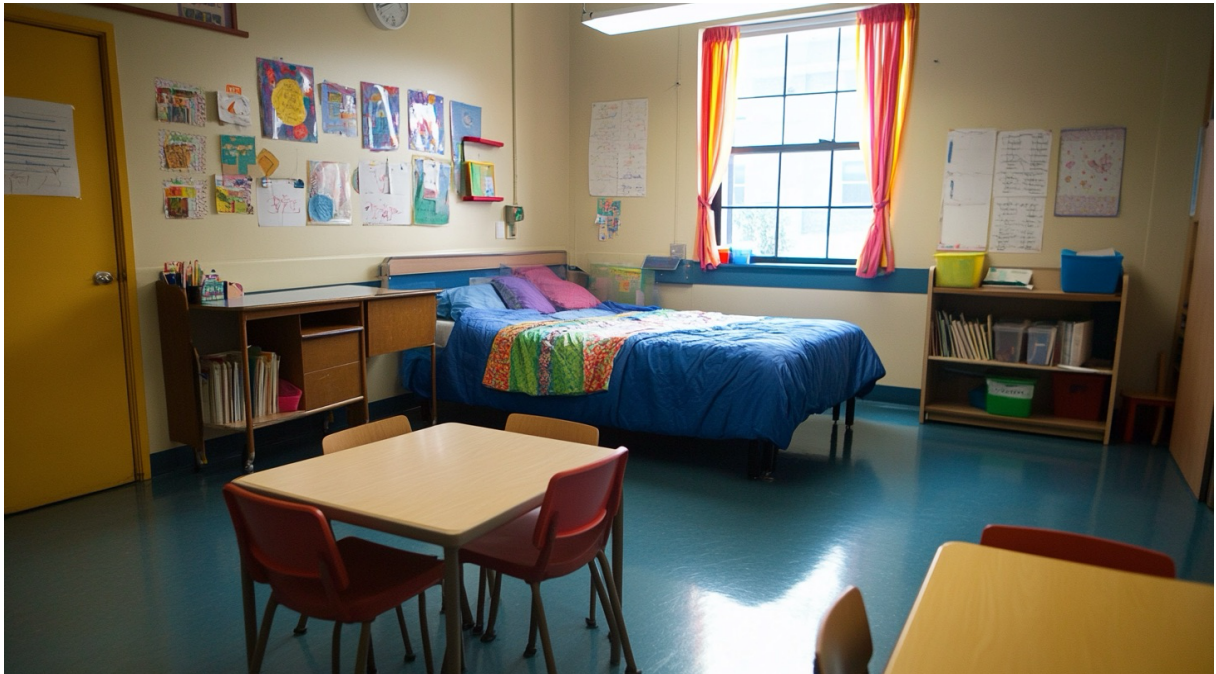

*Figure 2* AI-generated image created with Midjourney v.6.1 with the prompt "a school classroom with a hotel bed in the middle --ar 16:9 --style raw", 2024-11-20. Credit: author

For most image generators at time of writing, some statistical gradation between the provided words in the prompt and their associated visual material is created and this is most apparent in those prompts that pair disparate concepts or scenarios. In many cases, the model latches more strongly onto one concept over the others that appear in the prompt, which causes it to dominate the image output. In other cases, a more even mix is created. Another example created with the prompt "Trick-or-treaters wearing Christmas costumes," [Fig. 3] shows a more evenly mixed bleed between concepts and attendant imagery.

The setup and construction of this example hinges on the fact that children treat-or-treating on Halloween commonly wear certain types of costumes, such as pumpkins, skeletons, or monsters. They do not usually wear Christmas costumes, which have their own established tropes such as elves, Santa Claus, gingerbread men, and reindeer. The intention behind the prompt was that the depiction would show children on Halloween trick-or-treating in Christmas costumes such as these. The trappings of the activity (trick-or-treating) and the clothing (Christmas), as implied by the wording of the prompt, should be separate. As this image shows, however, a mix in themes has inevitably occurred. One of the children is wearing what appears to be a reindeer suit but, instead of brown fur, it has orange fur, which is not surprising given that the color orange dominates Halloween imagery. The children hold what appear to be classic Jack-O-Lantern pails for candy collecting, but the house and the door behind them is decorated for Christmas, with lights and a wreath on the door. A hybrid intermingling of the prompt concepts has thus been created and concretized. As these two images show, the seemingly specific and concrete elements of these image are also, at the same time, abstractions.

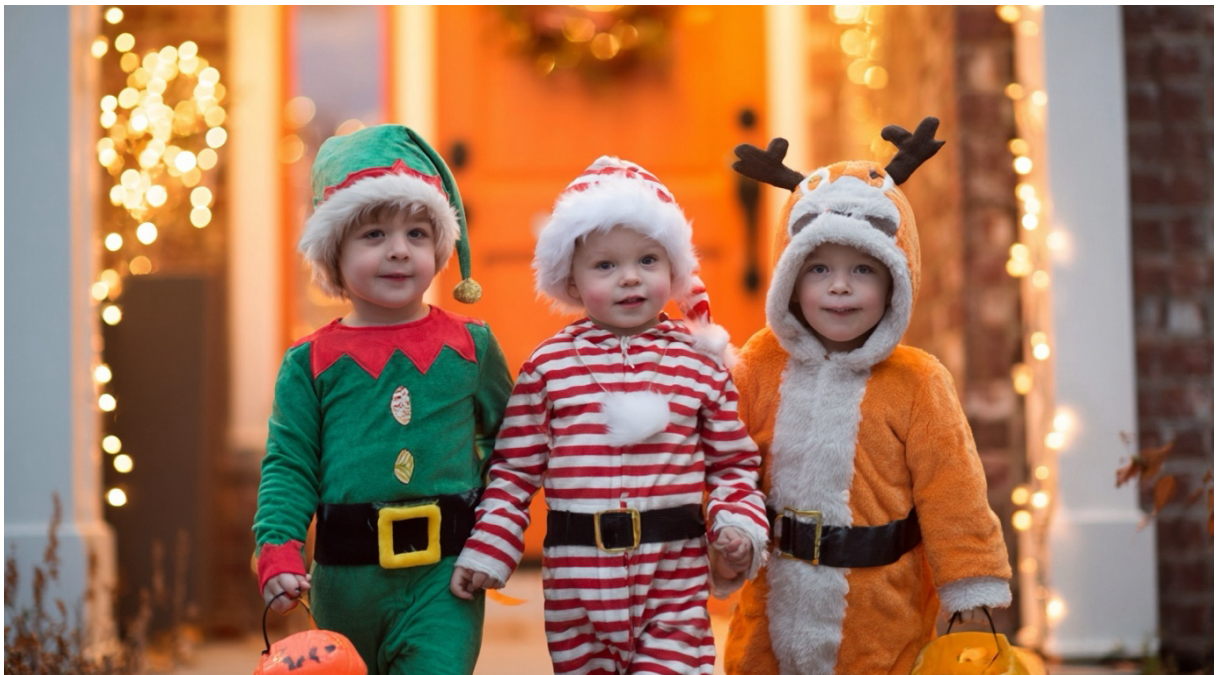

*Figure 3* AI-generated image created with Midjourney v.7.0 with the prompt "Trick-or-treaters wearing Christmas costumes --ar 16:9 –raw," 2025-08-30. Credit: author

**What do AI-generated pictures *really* want?**
The formulation of a spectrum from abstract to concrete (and back again) initially seems like a tidy way to describe representative form in service of understanding or conceptualizing the world. For AI-generated images, however, there is a vast disproportion in scale at play. In attempting to cross this gulf, they both achieve and fail to find concrete form. AI-generated images achieve concrete form in practical terms, given that a singular image can be produced. They fail, however, in the sense that no part of an AI-generated image can be considered separate from the statistical whole of collective representation.

The scale and unknowability of the origins of the concrete appearances of generative AI have spurred Silicon Valley into a frenzy of spiritualism around a prospective AI apocalypse, as described in the introduction to this article. The religious fervor over AI is echoed in the iconophilia and iconoclasms found with interdisciplinary debates over the use and nature of images. As art historian Caroline Jones eloquently writes, "One seeks the seamless thing in order to be astonished, all over again, by the magic of its madness. One breaks the spell in order to refuse the magic's clouding of the mind."[58] And it is precisely this element of generative AI that is most potent. The exchange of representational forms that generative AI models perform imitates the act of thought and the concept of the mind, which, in turn, effectively clouds the mind of the human interlocutor using these tools. As recent reports of the dire psychological effects of using AI chatbots indicate, generative AI acts on the mind of the user through representative forms: language and images.[59]

While this is certainly alarming, it is not the only—or, indeed, even the most pressing— existential threat to humanity posed by generative AI. Strikingly absent in the rhetoric of annihilation coming out of Silicon Valley is the threat that generative AI poses for the environment, climate change, and the use of resources on this planet. As is well-known at this point, data centers are enormously resource-draining, particularly in their use of water.

A single AI-generated image, like many of the incremental steps humanity takes toward climate-related self-destruction, is small in comparison to the vastness of the training data or the power it takes to train and run generative AI for a growing number of users. If, as the above discussion implies, AI-generated images want objecthood, materiality, or some basis in reality they have it in the reality of the resources consumed in their creation. What do AI-generated images *really* want, then? They want our water.

And, so, I cannot help but describe the desires of AI-generated images, as I have done above for Mitchell's pictures, as something that more resembles a horror film than a love story. The subsummation of pictures or the "collapse of distinction between pictures and persons" means that AI-generated images are treated as people—or, even, treated *better* than people—and, as such, they are given the potable water resources that should be going to human beings. In her in-depth history of Open AI, journalist Karen Hao describes how data centers have been built in some of the regions of the world with the scarcest water resources and given free rein to consume potable water while the population struggles to access clean water.[60] AI-generated images are not, therefore, the downtrodden and discriminated pictures-as-people that Mitchell compares to subjugated groups but, rather, they are kings and idols. As such, they are given our most valuable offerings.

Even so, the holy trinity that defines a picture for Mitchell (image, object, and discourse) remains unreachable for AI-generated images. Somewhere at their origin or in the process of their instantiation, resources are consumed but where and which ones exactly? A 2025 study investigating the water use of AI tools and the data centers associated with them estimated that GPT-3 had to use the equivalent of a 500ml bottle of water to produce 10 to 50 medium-length responses.[61] But even this is an abstraction. One cannot point to the actual water used in the making of a single picture or the specific data center/chip built to produce it or any other physical locus of their individual material origin. The complexity of generative AI—the whole system and apparatus created around it—and its physical and virtual manifestations make it so that even the materiality that underpins, produces, and inevitably propagates AI-generated images is abstract. The greedy desires of these images, therefore, are always at least partially hidden from view.

---

[58] Jones, "Making Abstraction," 416.
[59] Hamilton Morrin et al., "Delusions by Design? How Everyday AIs Might Be Fuelling Psychosis (and What Can Be Done about It)," preprint, OSF, July 11, 2025, https://doi.org/10.31234/osf.io/cmy7n_v5; Søren Dinesen Østergaard, "Will Generative Artificial Intelligence Chatbots Generate Delusions in Individuals Prone to Psychosis?," *Schizophrenia Bulletin* 49, no. 6 (2023): 1418–19, https://doi.org/10.1093/schbul/sbad128.
[60] Hao, *Empire of AI*, 272–300.
[61] Pengfei Li et al., "Making AI Less 'Thirsty,'" *Commun. ACM* 68, no. 7 (2025): 56, https://doi.org/10.1145/3724499.

       What is it, then, that AI-generated images want? There may be no need, after all, to think through this question via Mitchell's conceit because there is no need to project human-like subjecthood onto AI-generated images in order to uncover their desires. This would be giving them less power than they already have. Simply put, AI-generated images want the be valid, the way nature is valid. The only way to break the spell is to foreclose the idea of their inevitability.